\documentclass[10pt, twocolumn, conference]{IEEEtran}
\usepackage[utf8]{inputenc}

\usepackage[perpage,symbol]{footmisc}
\usepackage[english]{babel}
\usepackage{multirow,multicol}
\usepackage{amsmath, bm} 
\usepackage{pbox, booktabs, rotating}
\usepackage{subfigure}
\usepackage{array, makecell, lipsum}
\usepackage{enumerate}
\usepackage{threeparttable}
\usepackage{multirow,hyphenat,balance}
\usepackage{placeins}
\usepackage[compress]{cite}
\usepackage{float,color}
\usepackage{ragged2e}
\usepackage{enumerate}
\usepackage{etoolbox}
\usepackage[linesnumbered,ruled]{algorithm2e}
\usepackage{paralist}
\usepackage{cases}
\usepackage{url}
\usepackage{framed}
\usepackage{flushend}

\usepackage{amsthm}
\usepackage[english]{babel}

\theoremstyle{definition}

\newcolumntype{M}[1]{>{\centering\arraybackslash}m{#1}}

\usepackage[all=normal,paragraphs=tight,wordspacing=normal,charwidths=tight,indent=normal,floats=tight,leading=normal,lists=tight]{savetrees}

\newcommand*{\ie}{{i.e.}\@\xspace}

\newcommand*{\el}{et al.\@\xspace}

\title{ATPG-Guided Fault Injection Attacks on \\ Logic Locking}
\author{\IEEEauthorblockN{$^*$Ayush Jain, $^\dagger$M Tanjidur Rahman, and $^*$Ujjwal Guin \\}
\IEEEauthorblockA{$^*$ Dept. of Electrical and Computer Engineering, Auburn University\\ 
$^\dagger$ Dept. of Electrical and Computer Engineering, University of Florida \\
Emails: ayush.jain@auburn.edu, mir.rahman@ufl.edu, and ujjwal.guin@auburn.edu}
}

\begin{document}

\maketitle

\begin{abstract}
Logic Locking is a well-accepted protection technique to enable trust in the outsourced design and fabrication processes of integrated circuits (ICs) where the original design is modified by incorporating additional key gates in the netlist, resulting in a key-dependent functional circuit. The original functionality of the chip is recovered once it is programmed with the secret key, otherwise, it produces incorrect results for some input patterns. Over the past decade, different attacks have been proposed to break logic locking, simultaneously motivating researchers to develop more secure countermeasures. In this paper, we propose a novel stuck-at fault-based differential fault analysis (DFA) attack, which can be used to break logic locking that relies on a stored secret key. This proposed attack is based on self-referencing, where the secret key is determined by injecting faults in the key lines and comparing the response with its fault-free counterpart. A commercial ATPG tool can be used to generate test patterns that detect these faults, which will be used in DFA to determine the secret key. One test pattern is sufficient to determine one key bit, which results in at most \bm{$|K|$} test patterns to determine the entire secret key of size \bm{$|K|$}. %Stuck-at one \textit{sa1} faults at the key lines are selected while obtaining the test patterns. 
The proposed attack is generic and can be extended to break any logic locked circuits. %\todoAJ{Confirm with the VLSI Test book the notation for a stuck-at fault. Is it ``stuck-at one" or ``stuck-at-1"? {\color{blue} It is stuck-at-1 in the book } } %\todoAJ{Always put a comma before which.}
\end{abstract}

\vspace{5px}

\section{Introduction} \label{sec:intro}
Over the last few decades, the impact of globalization has transformed the semiconductor manufacturing and testing industry from vertical to horizontal integration. The continuous trend of device scaling has enabled the designer to incorporate more functionality in a system-on-chip~(SoC) by adopting lower technology nodes to increase performance and reduce the overall area and cost of an SoC. At present, majority of the SoC design companies or design houses no longer manufacture chips and maintain a foundry (fab) of their own due to cost for building and maintaining such foundries~\cite{YehFabCost2012} and the increased complexity in the fabrication process as new technology is adopted. The design-house integrates intellectual properties~(IP) obtained from different third-party IP vendors along with its design and outsources the manufacturing to an offshore foundry. Due to this distributed design and manufacturing flow, which includes third-party IPs, manufacturing, testing, and distribution of chips, various threats have emerged in recent years~\cite{alkabani2007active, castillo2007ipp, tehranipoor2011introduction}. The research community has also been extensively involved in proposing countermeasures against these threats~\cite{roy2008epic, rajendran2012security, charbon1998hierarchical, kahng2001constraint, qu2007intellectual, jarvis2007split}.

\begin{figure}[t]
    \centering
    \includegraphics[width=\linewidth]{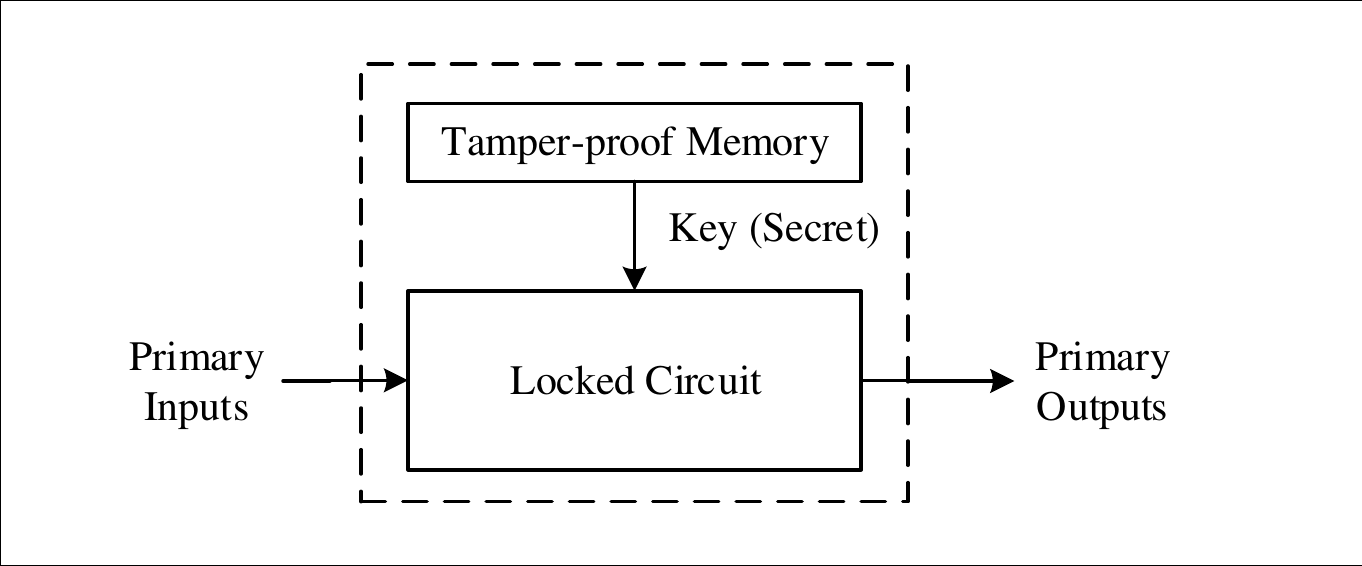} \vspace{-20px}
    \caption{An abstract view of the logic locking technique.} \vspace{-20px}
    \label{fig:LL}
\end{figure}
 
Logic locking has emerged as the most prominent method to address the threats incurred from untrusted manufacturing. In logic locking, the design of a circuit is locked so that the circuit produces incorrect results in normal operation unless a correct secret key is programmed into the chip. Figure~\ref{fig:LL} shows an abstract view of logic locking where the key is stored in a tamper-proof non-volatile memory. Subramanyan~\el~\cite{subramanyan2015evaluating} first showed that a locked circuit can efficiently be broken using key-pruning oracle-guided SAT analysis. Since then, many different versions of SAT-based attacks have been launched on logic locking~\cite{shamsi2019ip}, and the solutions have been proposed to mitigate these attacks as well~\cite{wang2017secure, sengupta2020truly, guin2016fortis, GuinVTS2017, GuinTVLSI2018, karmakar2018encrypt, potluriseql, chiang2019looplock, juretus2019increasing }.%\todoAJ{Any references from other groups??} 
\textit{Can we safely state that a logic locking technique is completely secure even if we achieve complete SAT resistivity?} Note that an untrusted foundry has many more effective means to determine the secret key without performing SAT analysis~\cite{rahmankey, jain2019taal, zhang2019tga, jainVTS2020}. Countermeasures are also developed to partially prevent these attacks~\cite{zhang2015veritrust, shen2018nanopyramid, vashistha2018trojan, zhang2019tga}.

%\todoAJ{Cite the 2020 VTS paper.}

In this paper, we show how an adversary can extract the secret key from a locked netlist, even if all the existing countermeasures are in place. An adversary can determine the secret key by injecting faults at the key registers~\cite{rahman2020defense, rahmankey}, which hold the key value during normal operation, and performing differential fault analysis. The entire process can be performed in three steps. First, an input pattern that reflects different output response for the change of only one key bit while keeping the other key bits at the faulty states is selected. To generate such a test pattern, we propose to use the constrained automatic test pattern generation (ATPG) algorithm, which is widely popular for testing VLSI circuits. The pattern which detects either stuck-at-1 (\textit{sa1}) or stuck-at-0 (\textit{sa0}) fault at one of the key lines with logic 1 or 0 constraints for other key lines respectively, is sufficient to determine that key bit. Note that the fault-free and faulty responses are always complementary for an input test pattern that detects that fault. The same process repeated for other key bits to obtain \bm{$|K|$} patterns for determining the entire key of size \bm{$|K|$}. Second, we apply these test patterns to two instances of an unlocked chip obtained from the market and collect the responses. Logic 1 faults are now injected into all the key lines and measure the responses of the faulty circuit by applying the set of test patterns. Next, logic 1 faults are injected into all the key lines except one key line. The response from this fault-free circuit is obtained by applying the test pattern associated with the fault-free key. This step is repeated for all the key lines, and responses are obtained by applying its corresponding test pattern. Finally, the results are compared to determine the key. The actual value of a key bit is 1 if the two responses are the same, otherwise, the key is 0. Note that this paper specifically considers the countermeasures in accordance with the logic locking techniques and not the countermeasures related to the prevention of fault injection or its detection.

The contributions of this paper are described as follows:

\begin{itemize}
    \item We propose a novel attack to break any key-based logic locking technique using the fault injection attack. When we apply a constrained \textit{sa1} pattern to a key line, the hypothesis key bit becomes 1 if the responses of the fault-free and faulty circuits are the same, otherwise, the key value is 0. The proposed attack is self-referencing and does not require any complex analysis (\ie, SAT). \textbf{\textit{To the best of our knowledge, we are the first to demonstrate that the stuck-at fault patterns can be used to determine the secret key of a locked circuit.}}
    \item We demonstrate and validate our proposed attack by performing the laser fault injection on a Kintex-7 FPGA. The technology-dependent gate-level netlist created in Synopsys Design Compiler is converted to a technology-independent netlist and implemented in Xilinx Vivado without any optimization so that the \textit{saf} patterns can be applied to the FPGA.
\end{itemize}

The rest of the paper is organized as follows: the proposed attack and its methodology to extract the secret key from any locked circuit are described in Section~\ref{sec:fault-injection-attack}. We present the results for the implementation of the proposed attack on different logic locked benchmark circuits in section~\ref{sec:experimental-results}. Finally, we conclude our paper in Section~\ref{sec:conclusion}.

\section{Proposed Fault Injection Attack}\label{sec:fault-injection-attack}
The differential fault analysis (DFA) attack on logic locking is motivated by the test pattern generation for VLSI circuits. A single stuck-at fault will be detected using a test pattern that activates the fault and propagates the faulty response to the primary output. The key register, which holds the key value loaded from the tamper-proof non-volatile memory, can be treated as the source of the fault. These registers are the target for an adversary to obtain the secret key from a working chip.  

\subsection{Threat Model}
The threat model defines the traits of an adversary and its position in the IC manufacturing and supply chain. It is very important to know an attacker's capabilities and its resources/tools to estimate its potential to launch the attack. The design house or entity designing the chip is assumed to be trusted. The attacker is assumed to be an untrusted foundry or a reverse engineer having access to the following:
\begin{itemize}
%\vspace{5px}
    \item The attacker has access to the locked netlist of a circuit. An untrusted foundry has access to all the layout information which can be extracted from the GDSII or OASIS file. Also, a locked netlist can be constructed from layer-by-layer reverse engineering of the fabricated chip with advanced technological tools~\cite{torrance2009state}. The attacker has the capability to determine the location of the tamper-proof memory. It can be trivial for an adversary to find the location of the key register in a netlist, as it can easily trace the route from the tamper-proof memory.
    \item The attacker has possession of an unlocked and fully functional chip, which can be easily acquired from the market.  
    \item A fault injection equipment is necessary to launch the attack. It is not necessary to use high-end fault injection equipment. The basic requirement is to inject faults at the key registers (all the flip-flops) location on a de-packaged/packaged chip. %Precise control is not necessary as we target all the flip-flops simultaneously.
\end{itemize}

\noindent\textbf{Notations}: An original circuit, and its locked version are denoted by $C_{O}$ and $C_L$, respectively. The two versions of fault-injected $C_L$ are represented as $C_{F}$ and $C_{A}$. $C_{F}$ represents a locked circuit where all the key lines ($|K|$) are injected with logic 1 (logic 0) faults, denoted as a faulty circuit. $C_{A}$ represents the same locked circuit where $(|K|-1)$ key lines are injected with the same logic 1 (logic 0) faults, leaving one key line fault free. We denote this circuit as a fault-free circuit for DFA. Both functional chips are loaded with the correct key in its tamper-proof memory. A fault is injected at the key register using a fault injection method (see details in Section~\ref{sec:experimental-results}). For any given circuit, we assume the primary inputs~($PI$) of size~\textit{$|PI|$}, primary outputs~($PO$) of size~\textit{$|PO|$}, and secret key~($K$) size of~\textit{$|K|$}. We also use key lines or key registers alternatively throughout this paper as their effects are the same on a circuit. Note that \textit{saf} is an abstract representation of a defect to generate test patterns, whereas, an injected fault is the manifestation of a faulty logic state due to fault injection.

\begin{figure}[ht]
    \centering
    \vspace{-5px}
    \includegraphics[width=\linewidth]{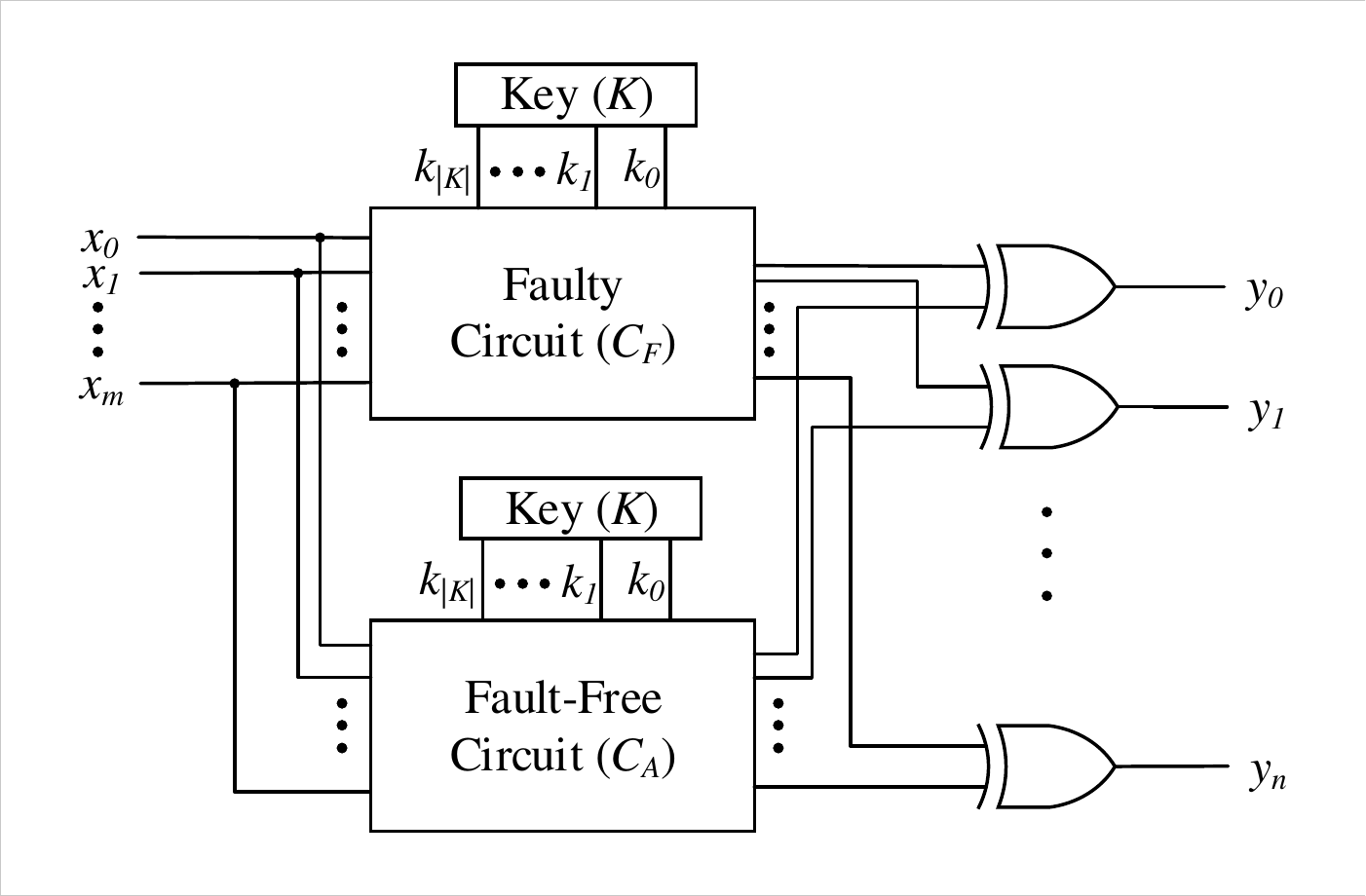} \vspace{-10px}
    \caption{The abstract representation of our proposed fault injection attack.} \vspace{-10px}
    \label{fig:absract-DFA}
\end{figure}

\subsection{Differential Fault Analysis Attack Methodology}\label{subsec:DFA}
The proposed fault injection attack relies on differential fault analysis, where the responses of two instances of faulty and fault-free circuits are compared to determine the secret key. A practical fault injection approach is described in Section~\ref{sec:experimental-results} to create the faulty chip. Figure~\ref{fig:absract-DFA} shows an abstract representation of our proposed approach. %The fault-free circuit ($C_A$) is an unlocked chip bought from the market whose key-bits needs to be retrieved. Except the targeted key-bit to be extracted, all remaining key registers are fixed to a particular faulty value of either 0 or 1 corresponding to the selected fault. While, faulty circuit ($C_F$) is the same chip, which is injected with a particular fault to keep all the key registers or interconnects to faulty value of logic 1 or 0. 
For an input pattern, the output responses are collected for both $C_A$ and $C_F$. The output responses are XORed to find any mismatch. If both the circuits differ in their responses, the XORed output will be 1, otherwise, it will be 0. If we find an input pattern that produces conflicting results for both $C_A$ and $C_F$ only for one key bit, the key value can be predicted. The key value is the same as the injected fault value if the XORed output is of logic 0, otherwise, the key value is complementary to the injected fault. The proposed attack can be described as follows:  

%\vspace{5px}
\noindent {\tiny $\bullet$} \textbf{\textit{Step-1}}: The first step is to select an input pattern that produces complementary results for the fault-free ($C_A$) and faulty ($C_F$) circuits. The input pattern needs to satisfy the following property --
it must sensitize only one key bit to the primary output(s). In other words, only the response of one key bit is visible at the \textit{PI} keeping all other key bits at logic 1s (or 0s). If this property is not satisfied, it will be impractical to reach a conclusion regarding the key bit value. Multiple key combinations can result in the same. \textit{Now the question is how can we find if such a pattern exists in the entire input space~($\xi$)}. 

To meet this requirement, our method relies on stuck-at faults~(\textit{saf}) based constrained ATPG to obtain the specific input test patterns (see details in Section~\ref{subsec:test-pattern-gen}). Considering the fact that adversary has access to the locked netlist~($C_L$), it can generate test patterns to detect \textit{sa1} or \textit{sa0} at any key lines and adding constraints to other key lines~(logic 1 and 0 for \textit{sa1} and \textit{sa0}, respectively). A single fault, either \textit{sa0} or \textit{sa1} on a key line is sufficient to determine the value of that key bit. Therefore, we have selected \textit{sa1} and the following sections are explained considering this fault only. This process is iterated over all the key-bits to obtain \bm{$|K|$} test patterns. The algorithm to generate the complete test pattern set is provided in Algorithm~section~\ref{subsec:test-pattern-gen}. 
    
%\vspace{5px}
\noindent {\tiny $\bullet$} \textbf{\textit{Step-2}}: The complete set of generated test patterns is applied to fault-induced functional circuit~($C_F$). The circuit is obtained by injecting logic 1 fault on the key registers if \textit{sa1} is selected in the previous step, else, logic 0 fault is injected for \textit{sa0}. The responses are collected for later comparison with the fault-free responses. For ($C_A$), the test patterns are applied such that it matches the fault modifications in the circuit. For example, the test pattern for the first key is applied to the circuit when the circuit instance does not pertain any fault on its corresponding key register and holds the correct key value while, the remaining key registers are set to logic 1 (for \textit{sa1}) or 0 (for \textit{sa0}). For the next key-bit, ($C_A$) instance is created by excluding this selected key bit from any fault while keeping all the other key registers to logic 1 (for \textit{sa1}) or 0 (for \textit{sa0}). This process is repeated for all key bits and their responses are collected for comparison in the next step. % { This process is iterated over all the instances of ($C_A$) by applying test pattern for the key-bit without any fault in the circuit. Not clear. }

%\vspace{5px}
\noindent {\tiny $\bullet$} \textbf{\textit{Step-3}}: The adversary will make the decision regarding the key value from the observed differences in the output responses of ($C_A$) and ($C_F$). For any test pattern corresponding to a particular key bit, when the output of both the circuits is same, it implies that the injected fault on the key lines in a $C_F$ circuit is same as the correct key bit, only then the output of both the ICs will be same. Otherwise, when $C_F$ and $C_A$ differ in their output response, it concludes the correct key bit is complementary to the induced fault. This process is repeated for all key bits. In this manner, the key value can be extracted by comparing the output responses of both circuits for the same primary input pattern.
  
\subsection{Example} \label{subsec:examples}

\begin{figure*}[ht]
    \centering
    \includegraphics[width=\linewidth]{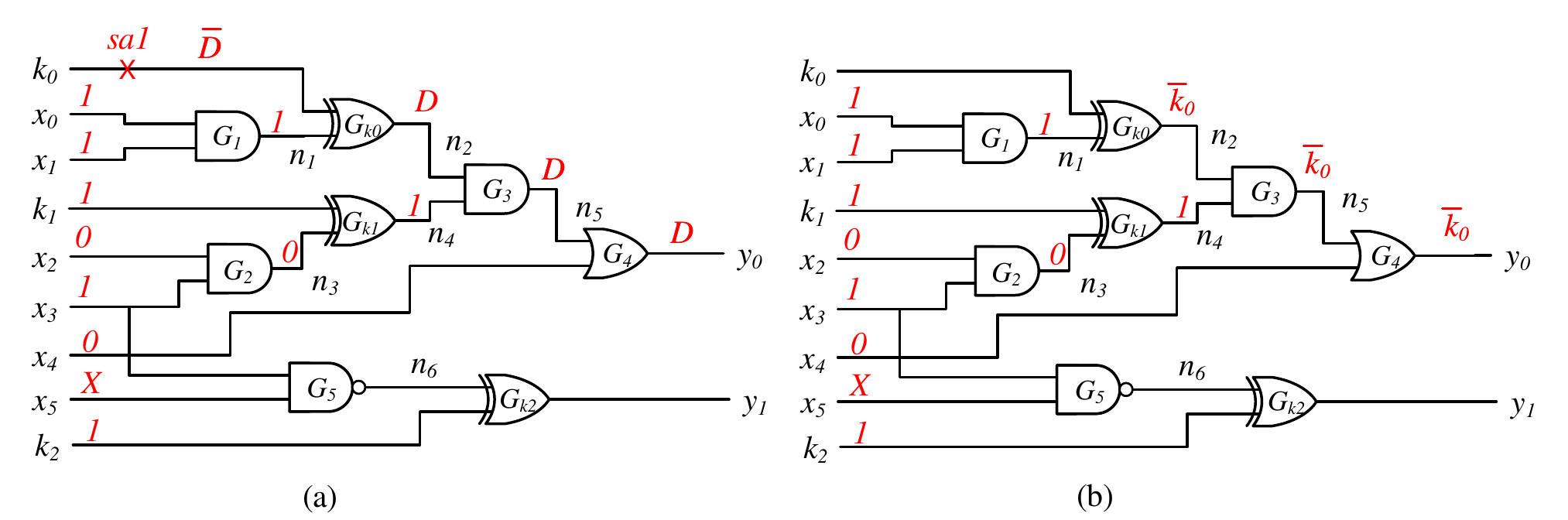} \vspace{-20px}
    \caption{Differential Fault attack on a test circuit locked with a 3-bit secret key, where the propagation of $k_0$ is dependent on $k_1$ and vice versa. (a) Test pattern generation considering a \textit{sa1} at key line $k_0$ with constraint $k_1=1$ and $k_2=1$. Test pattern, $P_1=[11010X]$ will be applied to $C_F$. (b) The same pattern are required to be applied to $C_A$. } 
    \label{fig:fault-propagation} \vspace{-15px}
\end{figure*}

In this section, we present an example circuit to illustrate the proposed attack. Test pattern generation for detecting stuck-at faults at the key lines is described using the D-Algorithm~\cite{bushnell2004essentials}. A Combinational circuit is chosen as an example for simplicity. However, the attack is valid for sequential circuits as well as it can be transformed into a combinational circuit in the scan mode, where all the internal flip-flops can be reached directly through the scan chains~\cite{bushnell2004essentials}.    %\todoAJ{Why are you referring FIAL??? Remove all instances of FIAL}

%Figure~\ref{fig:fault-propagation} shows a simple locked circuit with a 2-bit key, where the effect of one key does not impact the other key. The circuit has six inputs ($PI=6$) and two outputs ($PO=2$). It is necessary to generate a test pattern that detects a \textit{saf} at \textit{$k_0$} with constraint $k_1=1$, which is shown in Figure~\ref{fig:fault-propagation}.(a). $\overline{D}$ is assigned after the \textit{sa1} at the key line $k_0$. $D$ is defined as logic 1 for a good circuit and logic 0 for a faulty one~\cite{bushnell2004essentials}. To activate this fault, the ATPG tool will assign a logic 0 at $k_0$. It is required to propagate $\overline{D}$ to any of the primary outputs. For example, 1 at the output of gate $G_1$ will result in $D$ at the output of key gate $G_{k_0}$. Inputs $[x_0~x_1]=[1~1]$ can satisfy this condition since $G_1$ is an AND gate. Next, $D$ appears at $n_4$ for $[x_2~x_3]=[1~1]$. Finally to propagate $D$ at the output $y_0$, $x_4$ needs to be put to logic 0 as $G_4$ is an OR gate. As a result, $D$ is observed at the output $y_0$ for primary input $P_1=[x_0~x_1~x_2~x_3~x_4~x_5]$ equals to $[1~1~1~1~0~X]$. \textbf{\textit{Note that the output \bm{$y_0$} will have complementary values for \bm{$k_0=0$} and \bm{$k_0=1$} when we apply \bm{$P_1$} at the input.}} This property of the input patterns will be used in DFA to recover the secret key. Similar analysis can be performed to detect a \textit{saf} \textit{D} on key line $k_1$.

Figure~\ref{fig:fault-propagation} shows our proposed attack on a test circuit locked with a 3-bit secret key, where the propagation of $k_0$ and $k_1$ is inter-dependent on each other while propagation of $k_2$ is independent of other keys in the circuit. The circuit has six inputs ($|PI|=6$) and two outputs ($|PO|=2$). The attack targets all the key bits separately as mentioned before. First, we target to find out the value of $k_0$. It is necessary to generate a test pattern that detects a \textit{saf} at \textit{$k_0$} with constraint $k_1=1$ and $k_2=1$, which is shown in Figure~\ref{fig:fault-propagation}.(a). $\overline{D}$ is assigned after the \textit{sa1} at the key line $k_0$. $D$ is defined as logic 1 for a good circuit and logic 0 for a faulty one~\cite{bushnell2004essentials}. To activate this fault, the ATPG tool will assign a logic 0 at $k_0$. A test pattern $P_1$ needs to be generated to detect a \textit{sa1} fault at $k_0$ with constraint $k_1=1$ and $k_2=1$. As the value of $k_1$ is known during the pattern generation, the effect of the \textit{sa1} at $k_0$ will be propagated to the primary output $y_0$. For a fault value $\overline{D}$ at $k_0$, if $[x_0~x_1] = [1~1]$ then $D$ propagates to $n_2$ as $G_1$ is an AND gate. To propagate the value at $n_2$ to the output of $G_3$, its other input ($n_4$) needs to attain logic 1. Since $k_1=1$ due to injected fault which is set as a constraint in ATPG tool, $n_4=1$ for $n_3=0$ which implies $[x_2~x_3] = [0~1]$. At last, $x_4=0$ propagates $D$ value at $n_5$ to the primary output $y_0$. The output $y_0$ can be observed as $D$ for the test pattern $P_1=[1~1~0~1~0~X]$. \textbf{\textit{Note that the output \bm{$y_0$} will have complementary values for \bm{$k_0=0$} and \bm{$k_0=1$} when we apply \bm{$P_1$} at the input.}} This property of the input patterns will be used in DFA to recover the secret key. Similar analysis can be performed to detect \textit{saf} \textit{D} on other two key lines, $k_1$ and $k_2$.

After generating the test pattern $P_1$ for the \textit{sa1} at key line $k_0$, the next step is to perform differential fault analysis between the responses of the $C_F$ and $C_A$. The test pattern is applied first to the faulty circuit $C_F$ and its response is captured, which is shown in Figure~\ref{fig:fault-propagation}.(a). As this pattern detects a \textit{sa1} at line $k_0$, the faulty response will be propagated to the output $y_0$. If we injected a logic 1 fault ($D$) using the fault injection method, the value at $y_0$ will be logic 0 ($\overline{D}$). The same test pattern $P_1$ is now applied to the fault-free circuit $C_A$, which is shown in Figure~\ref{fig:fault-propagation}.(b). The logic value of $y_0$ for $C_A$ will be $\overline{k_0}$. If the value of $y_0$ is the same for both $C_F$ and $C_A$, the value of the key ($k_0$) is 1, otherwise, $k_0$ is equal to 0. Similarly, the test pattern for detecting a \textit{sa1} at $k_1$ can be applied to extract its value based on the difference between the two circuit instances.

\subsection{Test Pattern Generation} \label{subsec:test-pattern-gen}
To generate the test pattern set, an automated process relying on constrained ATPG is performed. The detailed steps to be followed are provided in Algorithm~\ref{alg:TP-generation}. Synopsys Design Compiler~\cite{SynopsysDC} is utilized to generate the technology-dependent gate-level netlist and its test protocol from the RTL design. A test protocol is required for specifying signals and initialization requirements associated with design rule checking in Synopsys TetraMAX~\cite{SynopsysTetraMAX}. Automatic test generation tool TetraMAX generates the test patterns for the respective faults along with constraints for the locked gate-level netlist.   

\vspace{-5px}
\begin{algorithm}[]
\SetAlgoLined
\SetKwInOut{Input}{Input}\SetKwInOut{Output}{Output}
\Input{~Locked gate-level netlist~($C_L$), test protocol~($T$), and standard cell library}
\Output{~Test pattern (\textit{P}) set}

% \vspace{5px}
% \hline
% \vspace{5px}

Read the locked netlist ($C_L$) \;
Read standard cell library \;

Run design rule check with test protocol generated from design compiler \;

Determine key size~\bm{$|K|$} from $C_L$ \; 

\For{$i\gets0$ \KwTo ($|K|-1$) }{
    Add a \textit{sa1} fault at key line $k_i$ \;
        
        \For{$j\gets0$ \KwTo ($|K|-1$)}{
            \If{$i \neq j$}{
                Add constraint at $k_j$ to logic 1 \;
            }
        }
        Run ATPG to detect the fault \;
        Add the test pattern, $P_i$ to the pattern set, \textit{P} \;
        Remove all faults; Remove all constraints \;
    }
 Report the test pattern set, $P$ \;
\caption{Test pattern generation for constrained ATPG} \label{alg:TP-generation}
\end{algorithm} %\todoAJ{Be consistent with small and large caps.}

\vspace{-5px}

The inputs to the algorithm are locked gate-level netlist~($C_L$), Design Compiler generated test protocol~($T$), and the standard cell library. The algorithm starts with reading the locked netlist and standard cell library~(Lines 1-2). The ATPG tool runs the design rule check with the test protocol obtained from the Design Compiler to check for any violation (Line 3). Only upon completion of this step, the fault model environment is set up in the tool. The size of the key ($|K|$) is determined by analyzing $C_L$ (Line 4). The remaining key lines are selected one by one to generate test patterns (Line 5). A stuck-at-1 fault is added at the $i^{th}$ key line to generate $P_i$ (Line 6). The ATPG constraints (logic 1) are added to other key lines (Lines 7-10). A test pattern $P_i$ is generated to detect the \textit{sa1} at the $i^{th}$ key line (Lines 12-13) and added to the pattern set, $P$. All the added constrains and faults are removed to generate the $(i+1)^{th}$ test pattern (Line 14). Finally, the algorithm reports all the test patterns, $P$ (Line 16).

\subsection{Fault-injection Approach}
Fault-injection attack has been widely used in the past to extract secret assets and bypassing security measures in the device~\cite{kim2007faults}. An adversary can use several fault-injection approaches depending on the budget and expertise. The basic fault-injection approach includes voltage, timing, electromagnetic, and laser-based fault-injection methods~\cite{moro2013electromagnetic, alam2019ram, tajik2015laser}. Laser-fault injection~(LFI) offers the most precision in both spatial and temporal domains during the operation of the chip, hence, used for deploying DFA attack for extracting the secret key. Laser with photon energy higher than silicon bandgap energy used to induce faults in an integrated circuit~\cite{tajik2015laser}. Therefore, the laser with a wavelength less than 1.1~$\mu$m is used in our experiment. The LFI attack can be completed in the following steps: 

%\vspace{5px}
\noindent {\tiny $\bullet$} \textbf{Sample Preparation:}  LFI can be injected from both frontside and backside of the chip. However, the interconnecting metal layers at the front of the die obstruct the optical path of photons. On the other hand, the absence of any metal obstacle or reflective coating at the backside of the die allows an adversary to access the transistors with the laser. In a typical packaged chip (bondwire IC), the backside can be exposed by wet etching. Nonetheless, the flip-chip substrate is typically covered with a metallic lid, which can be easily removed to expose the silicon die. The backside of the silicon can be further polished to 30 -- 100~$\mu$m to reduce the power loss along the laser path due to photon absorption phenomena ~\cite{champeix2015seu, tajik2015laser}.   

%\vspace{5px}
\noindent {\tiny $\bullet$} \textbf{Target Localization and Fault-injection:} The method of localizing key-register location depends on the capability and asset availability to an adversary. An adversary, like an untrusted foundry or an expert reverse engineer, can localize the key location, i.e., tamper-proof memory, key-register, key-gates by analyzing the GDSII or partial/full-blown reverse engineering. Once the target is localized, an attacker needs to identify the fault sensitive location for injecting fault. Localizing the most reverse biased P-N junction in the key-register can be identified as the potential candidate for fault-injection~\cite{champeix2015seu}. Therefore, depending on logic 1 (logic 0) fault, the laser can be applied to the drain location of the p-type (n-type) MOS transistors for fault injection. 

Another challenge is that a single laser source can only inject a single fault at once. Therefore, the fault can be injected in a sequential order where the laser source can be moved from one key-register to another for injecting fault. After localizing the targeted key registers, an adversary can automate the sequential fault-injection process with the help of computer vision and image processing~\cite{stellari2018automated, vashistha2018trojan}. Since the key is imperative for the IP operation, it is safe to assume that once secure boot-up is complete, the locking key will remain stored in the key-register during the operation of IP~\cite{rahman2020defense, rahmankey}. Therefore, an adversary can initiate the fault-injection method after the secure boot-up of the chip is complete. An adversary can identify the clock-cycle required for secure-boot up by monitoring the power consumption of the circuit. 

\section{Experimental Results} \label{sec:experimental-results}
To evaluate the effectiveness of our proposed attack, we adopted and performed the laser fault injection technique on a Kintex-7 FPGA, which is used as the device-under-test (DUT). Different benchmark circuits are implemented in a Kintex-7 FPGA, where the faults are injected on the key registers. First, the RTL netlist for ISCAS'99 benchmark circuits~\cite{bryan1985iscas} are synthesized using 32nm technology libraries in Synopsys Design Compiler~\cite{SynopsysDC}. The technology-dependent gate-level locked netlist is given to the Synopsys TetraMAX ATPG tool~\cite{SynopsysTetraMAX} to generate test pattern set~\textit{P} using Algorithm~\ref{alg:TP-generation}. The same netlist is then converted into a technology-independent gate-level Verilog code using our in-house PERL script. This is primarily done to assure that the circuit implemented in the FPGA is exactly the same circuit for which the test pattern set is generated. Otherwise, fault propagation cannot be ensured. Fault injection is performed on the circuit loaded into the FPGA, which leads to the instances of faulty and fault-free circuits by laser-induced faults on the key registers. Additionally, the implemented design includes a separate universal asynchronous receiver/transmitter~(UART) module, which is used for communication between the computer and the FPGA. The inputs are applied through the real-term monitor and responses are collected on the same. Once the response for any key-bit is obtained, the step is repeated for all the key bits in a benchmark circuit. Finally, the key-bits are exposed through the comparison between the corresponding instances of the circuits as explained in Section~\ref{subsec:DFA}. 
\begin{figure}[h]
	%\vspace{-0.25pt}
        \centering
        \includegraphics[width = 0.9\linewidth]{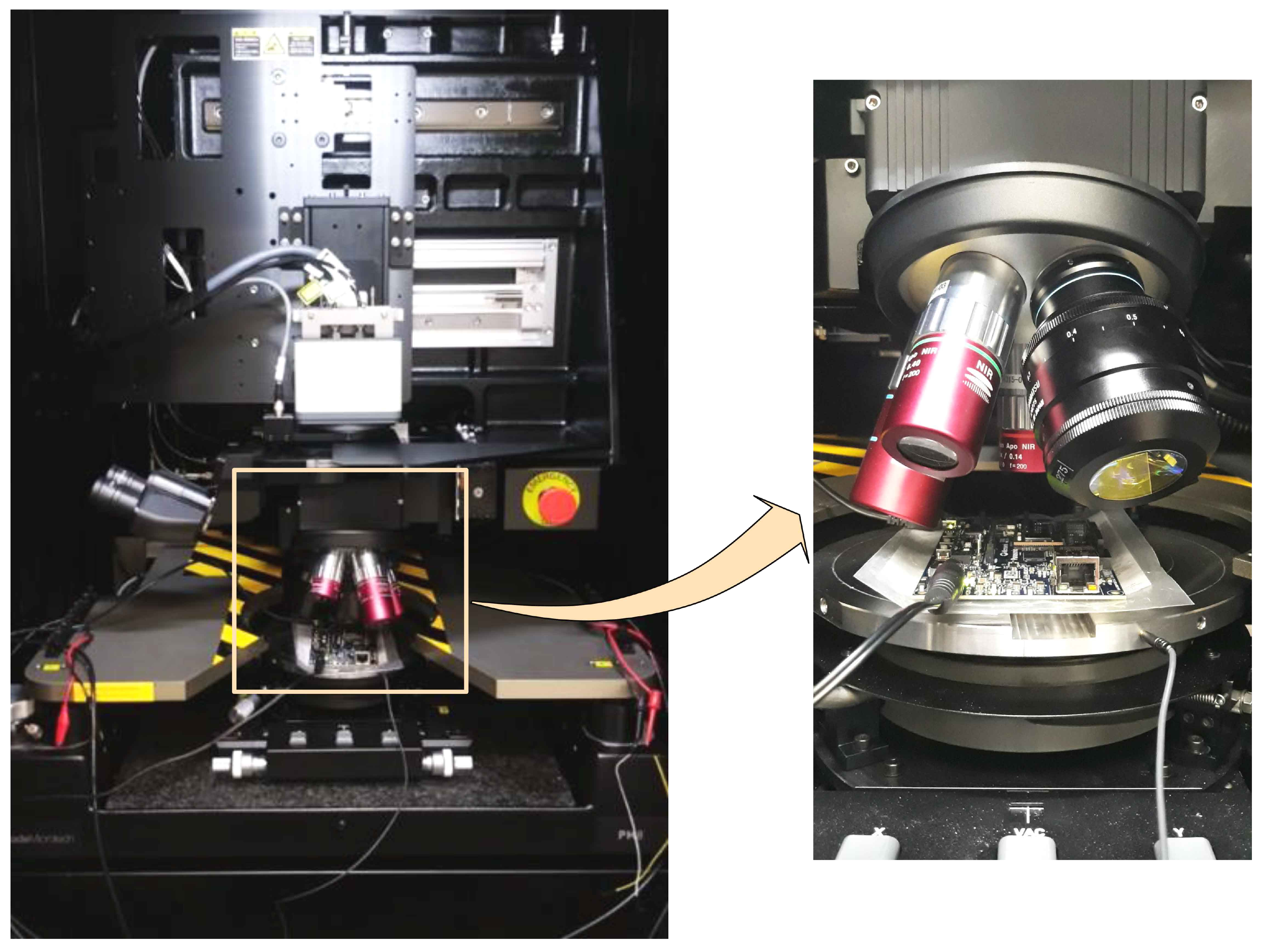} \vspace{-5px}
        \caption{The FPGA board placed under the lens for laser-fault injection at the target registers.}
        \vspace{-15px}
        \label{fig:setup}
\end{figure}

%\todoAJ{Can you add a zoomed center piece, beside this pic and put an arrow? }

\subsection{Laser Fault Injection Attack}\label{subsec:laser-fault-injection}
The laser fault injection~(LFI) setup is provided by a Hamamatsu PHEMOS-1000 FA microscope as shown in~\ref{fig:setup}. The equipment consists of a diode pulse laser source (Hamamatsu C9215-06) with a wavelength of 1064 $nm$. Three objective lenses were used during this work: 5x/0:14 NA, 20x/0:4 NA, 50x/0:76 NA. The 50x lens is equipped with a correction ring for silicon substrate thickness. The laser diode have two operation modes -- a) low power (200 $mW$) pulse mode, and b) high power (800 $mW$) impulse mode. The high power impulse mode can be used for laser fault injection. The laser power can be adjusted from 2$\%$ to 100$\%$ in 0.5$\%$ steps. 

Photon emission analysis~\cite{rahman2019backside} is used to localize the implemented locked circuitry in the DUT. Thereafter, The DUT is placed under the laser source for LFI. A trigger signal is fed to the PHEMOS-1000 to synchronize the LFI with DUT operation. Once the device reaches a stable state after power-on, the laser is triggered on target key-registers. After the fault injection, we have to guarantee that the device is still functioning as expected and has not entered into a completely dysfunctional state. The laser triggering timing can be checked by a digital oscilloscope for greater precision. 

We have performed and verified our results for different benchmark circuits implemented with random logic locking (RLL)~\cite{roy2010ending}, strong interference-based logic locking~(SLL)~\cite{rajendran2012security} and fault-based stripped functionality logic locking~(SFLL-Fault)~\cite{sengupta2018customized}. For RLL, we selected locked instances of c432 and c2670 benchmark circuits with a 32-bit key and 128-bit key respectively obtained from Trust-hub~\cite{salmani2018trust}. For SLL, we selected c1355 and c1908 locked benchmarks with 128-key bits, also obtained from Trust-Hub. We also implemented the attack on the circuit locked with a combination of SFLL-fault~(40-bit key) and RLL~(40-bit key) technique. We successfully recovered the entire key for all the circuits which proves the effectiveness of our proposed ATPG-guided fault injection attack.

\section{Conclusion}\label{sec:conclusion}
In this paper, we have presented a novel ATPG-guided stuck-at fault based attack to undermine the security of any logic locking technique. The attack relies on injecting faults on the key lines through hardware to perform differential fault analysis between faulty and fault-free chip for the ATPG generated test patterns. We have shown that at most \bm{$|K|$} test patterns are required to recover the entire secret key of size \bm{$|K|$}. We have demonstrated the attack on circuits implemented in the FPGA using the laser fault injection method. The results depicted the success of the proposed attack on different logic locking techniques, irrespective of their SAT resiliency.
%\todoAJ{Reduce 1/3 texts in the conclusion}

\section*{Acknowledgment}
The authors would like to thank Dr. Navid Asadizanjani, University of Florida, for helping with laser fault injection experimentation. This work was supported in part by the National Science Foundation under grant number CNS-1755733 and Air Force Research Laboratory under grant AF-FA8650-19-1-1707.

%\newpage

\bibliographystyle{IEEEtran}
\bibliography{bib-FI}

\begin{thebibliography}{10}
\providecommand{\url}[1]{#1}
\csname url@samestyle\endcsname
\providecommand{\newblock}{\relax}
\providecommand{\bibinfo}[2]{#2}
\providecommand{\BIBentrySTDinterwordspacing}{\spaceskip=0pt\relax}
\providecommand{\BIBentryALTinterwordstretchfactor}{4}
\providecommand{\BIBentryALTinterwordspacing}{\spaceskip=\fontdimen2\font plus
\BIBentryALTinterwordstretchfactor\fontdimen3\font minus
  \fontdimen4\font\relax}
\providecommand{\BIBforeignlanguage}[2]{{%
\expandafter\ifx\csname l@#1\endcsname\relax
\typeout{** WARNING: IEEEtran.bst: No hyphenation pattern has been}%
\typeout{** loaded for the language `#1'. Using the pattern for}%
\typeout{** the default language instead.}%
\else
\language=\csname l@#1\endcsname
\fi
#2}}
\providecommand{\BIBdecl}{\relax}
\BIBdecl

\bibitem{YehFabCost2012}
{Age Yeh}, ``{Trends in the global IC design service market},'' DIGITIMES
  Research, 2012.

\bibitem{alkabani2007active}
Y.~Alkabani and F.~Koushanfar, ``{Active Hardware Metering for Intellectual
  Property Protection and Security.}'' in \emph{USENIX security symposium},
  2007, pp. 291--306.

\bibitem{castillo2007ipp}
E.~Castillo, U.~Meyer-Baese, A.~Garc{\'\i}a, L.~Parrilla, and A.~Lloris,
  ``{IPP@ HDL: efficient intellectual property protection scheme for IP
  cores},'' \emph{IEEE Trans. on Very Large Scale Integration (VLSI) Systems},
  pp. 578--591, 2007.

\bibitem{tehranipoor2011introduction}
M.~Tehranipoor and C.~Wang, \emph{Introduction to hardware security and
  trust}.\hskip 1em plus 0.5em minus 0.4em\relax Springer Science \& Business
  Media, 2011.

\bibitem{roy2008epic}
J.~A. Roy, F.~Koushanfar, and I.~L. Markov, ``{EPIC: Ending piracy of
  integrated circuits},'' in \emph{Proceedings of the conference on Design,
  automation and test in Europe}, 2008, pp. 1069--1074.

\bibitem{rajendran2012security}
J.~Rajendran, Y.~Pino, O.~Sinanoglu, and R.~Karri, ``Security analysis of logic
  obfuscation,'' in \emph{Proc. of Annual Design Automation Conference}, 2012,
  pp. 83--89.

\bibitem{charbon1998hierarchical}
E.~Charbon, ``{Hierarchical watermarking in IC design},'' in \emph{Proc. of the
  IEEE Custom Integrated Circuits Conference}, 1998, pp. 295--298.

\bibitem{kahng2001constraint}
A.~B. Kahng, J.~Lach, W.~H. Mangione-Smith, S.~Mantik, I.~L. Markov,
  M.~Potkonjak, P.~Tucker, H.~Wang, and G.~Wolfe, ``{Constraint-based
  watermarking techniques for design IP protection},'' \emph{IEEE Transactions
  on Computer-Aided Design of Integrated Circuits and Systems}, pp. 1236--1252,
  2001.

\bibitem{qu2007intellectual}
G.~Qu and M.~Potkonjak, \emph{{Intellectual property protection in VLSI
  designs: theory and practice}}.\hskip 1em plus 0.5em minus 0.4em\relax
  Springer Science \& Business Media, 2007.

\bibitem{jarvis2007split}
R.~W. Jarvis and M.~G. McIntyre, ``Split manufacturing method for advanced
  semiconductor circuits,'' 2007, uS Patent 7,195,931.

\bibitem{subramanyan2015evaluating}
P.~Subramanyan, S.~Ray, and S.~Malik, ``Evaluating the security of logic
  encryption algorithms,'' in \emph{IEEE International Symposium on Hardware
  Oriented Security and Trust (HOST)}, 2015, pp. 137--143.

\bibitem{shamsi2019ip}
K.~Shamsi, M.~Li, K.~Plaks, S.~Fazzari, D.~Z. Pan, and Y.~Jin, ``{IP Protection
  and Supply Chain Security through Logic Obfuscation: A Systematic
  Overview},'' \emph{Trans. on Design Automation of Electronic Systems
  (TODAES)}, p.~65, 2019.

\bibitem{wang2017secure}
X.~{Wang}, D.~{Zhang}, M.~{He}, D.~{Su}, and M.~{Tehranipoor}, ``{Secure Scan
  and Test Using Obfuscation Throughout Supply Chain},'' \emph{IEEE
  Transactions on Computer-Aided Design of Integrated Circuits and Systems},
  vol.~37, no.~9, pp. 1867--1880, Sep. 2018.

\bibitem{sengupta2020truly}
A.~Sengupta, M.~Nabeel, N.~Limaye, M.~Ashraf, and O.~Sinanoglu, ``Truly
  stripping functionality for logic locking: A fault-based perspective,''
  \emph{Transactions on Computer-Aided Design of Integrated Circuits and
  Systems}, 2020.

\bibitem{guin2016fortis}
U.~Guin, Q.~Shi, D.~Forte, and M.~M. Tehranipoor, ``{FORTIS: a comprehensive
  solution for establishing forward trust for protecting IPs and ICs},''
  \emph{ACM Transactions on Design Automation of Electronic Systems (TODAES)},
  p.~63, 2016.

\bibitem{GuinVTS2017}
U.~Guin, Z.~Zhou, and A.~Singh, ``{A novel design-for-security (DFS)
  architecture to prevent unauthorized IC overproduction},'' in \emph{Proc. of
  the IEEE VLSI Test Symposium (VTS)}, 2017, pp. 1--6.

\bibitem{GuinTVLSI2018}
------, ``{Robust design-for-security architecture for enabling trust in IC
  manufacturing and test},'' \emph{Trans. on Very Large Scale Integration
  (VLSI) Systems}, pp. 818--830, 2018.

\bibitem{karmakar2018encrypt}
R.~Karmakar, S.~Chatopadhyay, and R.~Kapur, ``{Encrypt flip-flop: A novel logic
  encryption technique for sequential circuits},'' \emph{arXiv preprint
  arXiv:1801.04961}, 2018.

\bibitem{potluriseql}
S.~Potluri, A.~Kumar, and A.~Aysu, ``{SeqL: SAT-attack Resilient Sequential
  Locking},'' 2020.

\bibitem{chiang2019looplock}
H.-Y. Chiang, Y.-C. Chen, D.-X. Ji, X.-M. Yang, C.-C. Lin, and C.-Y. Wang,
  ``{LOOPLock: LOgic OPtimization based Cyclic Logic Locking},''
  \emph{Transactions on Computer-Aided Design of Integrated Circuits and
  Systems}, 2019.

\bibitem{juretus2019increasing}
K.~Juretus and I.~Savidis, ``{Increasing the SAT Attack Resiliency of In-Cone
  Logic Locking},'' in \emph{International Symposium on Circuits and Systems
  (ISCAS)}, 2019, pp. 1--5.

\bibitem{rahmankey}
M.~T. Rahman, S.~Tajik, M.~S. Rahman, M.~Tehranipoor, and N.~Asadizanjani,
  ``The key is left under the mat: On the inappropriate security assumption of
  logic locking schemes,'' in \emph{Conference on IEEE Int. Sym. on Hardware
  Oriented Security and Trust (HOST)}, 2020.

\bibitem{jain2019taal}
A.~Jain, Z.~Zhou, and U.~Guin, ``{TAAL: Tampering Attack on Any Key-based Logic
  Locked Circuits},'' \emph{arXiv preprint arXiv:1909.07426}, 2019.

\bibitem{zhang2019tga}
Y.~Zhang, P.~Cui, Z.~Zhou, and U.~Guin, ``{TGA: An Oracle-less and
  Topology-Guided Attack on Logic Locking},'' in \emph{Proceedings of the 3rd
  ACM Workshop on Attacks and Solutions in Hardware Security Workshop}, 2019,
  pp. 75--83.

\bibitem{jainVTS2020}
A.~Jain, U.~Guin, M.~T. Rahman, N.~Asadizanjani, D.~Duvalsaint, and R.~D.~S.
  Blanton, ``{Special Session: Novel Attacks on Logic-Locking},'' in \emph{VLSI
  Test Symposium (VTS)}, 2020.

\bibitem{zhang2015veritrust}
J.~Zhang, F.~Yuan, L.~Wei, Y.~Liu, and Q.~Xu, ``{VeriTrust: Verification for
  hardware trust},'' \emph{Transactions on Computer-Aided Design of Integrated
  Circuits and Systems}, pp. 1148--1161, 2015.

\bibitem{shen2018nanopyramid}
H.~Shen, N.~Asadizanjani, M.~Tehranipoor, and D.~Forte, ``{Nanopyramid: An
  Optical Scrambler Against Backside Probing Attacks},'' in \emph{Proc. Int.
  Symposium for Testing and Failure Analysis(ISTFA)}, 2018, p. 280.

\bibitem{vashistha2018trojan}
N.~Vashistha, H.~Lu, Q.~Shi, M.~T. Rahman, H.~Shen, D.~L. Woodard,
  N.~Asadizanjani, and M.~Tehranipoor, ``{Trojan scanner: Detecting hardware
  trojans with rapid SEM imaging combined with image processing and machine
  learning},'' in \emph{Proceedings from the International Symposium for
  Testing and Failure Analysis (ISTFA)}.\hskip 1em plus 0.5em minus 0.4em\relax
  ASM International, 2018, p. 256.

\bibitem{rahman2020defense}
M.~T. Rahman, M.~S. Rahman, H.~Wang, S.~Tajik, W.~Khalil, F.~Farahmandi,
  D.~Forte, N.~Asadizanjani, and M.~Tehranipoor, ``Defense-in-depth: A recipe
  for logic locking to prevail,'' \emph{Integration}, 2020.

\bibitem{torrance2009state}
R.~Torrance and D.~James, ``{The state-of-the-art in IC reverse engineering},''
  in \emph{International Workshop on Cryptographic Hardware and Embedded
  Systems}, 2009, pp. 363--381.

\bibitem{bushnell2004essentials}
M.~Bushnell and V.~Agrawal, \emph{{Essentials of electronic testing for
  digital, memory and mixed-signal VLSI circuits}}.\hskip 1em plus 0.5em minus
  0.4em\relax Springer Science \& Business Media, 2004, vol.~17.

\bibitem{SynopsysDC}
``{Synopsys Design Compiler},'' {Synopsys, Inc., 2017}.

\bibitem{SynopsysTetraMAX}
``{TetraMAX ATPG: Automatic Test Pattern Generation},'' synopsys, Inc., 2017.

\bibitem{kim2007faults}
C.~H. Kim and J.-J. Quisquater, ``Faults, injection methods, and fault
  attacks,'' \emph{Design \& Test of Computers}, vol.~24, no.~6, pp. 544--545,
  2007.

\bibitem{moro2013electromagnetic}
N.~Moro, A.~Dehbaoui, K.~Heydemann, B.~Robisson, and E.~Encrenaz,
  ``Electromagnetic fault injection: towards a fault model on a 32-bit
  microcontroller,'' in \emph{Workshop on Fault Diagnosis and Tolerance in
  Cryptography}.\hskip 1em plus 0.5em minus 0.4em\relax IEEE, 2013, pp. 77--88.

\bibitem{alam2019ram}
M.~M. Alam, S.~Tajik, F.~Ganji, M.~Tehranipoor, and D.~Forte, ``Ram-jam: Remote
  temperature and voltage fault attack on fpgas using memory collisions,'' in
  \emph{Workshop on Fault Diagnosis and Tolerance in Cryptography (FDTC)},
  2019, pp. 48--55.

\bibitem{tajik2015laser}
S.~Tajik, H.~Lohrke, F.~Ganji, J.-P. Seifert, and C.~Boit, ``Laser fault attack
  on physically unclonable functions,'' in \emph{Workshop on fault diagnosis
  and tolerance in cryptography (FDTC)}, 2015, pp. 85--96.

\bibitem{champeix2015seu}
C.~Champeix, N.~Borrel, J.-M. Dutertre, B.~Robisson, M.~Lisart, and
  A.~Sarafianos, ``Seu sensitivity and modeling using pico-second pulsed laser
  stimulation of a d flip-flop in 40 nm cmos technology,'' in
  \emph{International symposium on defect and fault tolerance in VLSI and
  nanotechnology systems (DFTS)}, 2015, pp. 177--182.

\bibitem{stellari2018automated}
F.~Stellari, C.-C. Lin, T.~Wassick, T.~Shaw, and P.~Song, ``Automated
  contactless defect analysis technique using computer vision,'' in
  \emph{Proceedings from the International Symposium for Testing and Failure
  Analysis (ISTFA)}, 2018, p.~79.

\bibitem{bryan1985iscas}
D.~Bryan, ``{The ISCAS'85 benchmark circuits and netlist format},'' \emph{North
  Carolina State University}, p.~39, 1985.

\bibitem{rahman2019backside}
M.~T. Rahman and N.~Asadizanjani, ``Backside security assessment of modern
  socs,'' in \emph{International Workshop on Microprocessor/SoC Test, Security
  and Verification (MTV)}, 2019, pp. 18--24.

\bibitem{roy2010ending}
J.~A. Roy, F.~Koushanfar, and I.~L. Markov, ``{Ending piracy of integrated
  circuits},'' \emph{Computer}, pp. 30--38, 2010.

\bibitem{sengupta2018customized}
A.~Sengupta, M.~Ashraf, M.~Nabeel, and O.~Sinanoglu, ``{Customized locking of
  IP blocks on a multi-million-gate SoC},'' in \emph{Int. Conf. on
  Computer-Aided Design (ICCAD)}, 2018, pp. 1--7.

\bibitem{salmani2018trust}
H.~Salmani and M.~Tehranipoor, ``{Trust-hub},'' 2018, [Online]. Available:
  https://trust-hub.org/home.

\end{thebibliography}

\end{document}